\documentclass[hyper]{JHEP} 

\usepackage{epsfig}

\newcommand\fverb{\setbox\pippobox=\hbox\bgroup\verb}
\newcommand\fverbdo{\egroup\medskip\noindent%
			\fbox{\unhbox\pippobox}\ }
\newcommand\fverbit{\egroup\item[\fbox{\unhbox\pippobox}]}
\newbox\pippobox
\title{Some Solutions of Berkovits'Superstring Field Theory}

\author{ J. Kluso\v{n}
\footnote{On leave from Masaryk University, Brno}\\
Institute of Theoretical Physics, University of Stockholm, SCFAB\\
SE- 106 91 Stockholm, Sweden \\
and \\
Institutionen f\"or teoretisk fysik\\
BOX 803, SE- 751 08 
Uppsala, Sweden \\
E-mail: \email{josef.kluson@teorfys.uu.se}}

\preprint{\hepth{0201054}}

\abstract{In this  note we construct
exact solution of 
Berkovits'superstring field theory.}

\begin{document}
\section{Introduction}\label{firstx}
Open string field theory is very useful tool 
for the understanding of the tachyon 
condensation.   (For  review  and
extensive list of references, see
\cite{Ohmori}).  While 
  bosonic
string field  theory \cite{WittenSFT}
 based on the Chern-Simons-like
 action is very well understood 
  in superstring case
the situation is not such a clear. The straightforward
superstring extension 
of the bosonic string field theory
 \cite{WittenSFT1} suffers from an emergence
of contact divergences  that
forces us to include higher-contact
 terms when two picture changing
operators collide. (For discussion of 
this issue and other related problems we
again recommend to see very good review
\cite{Ohmori}.)

Another interesting approach to the problem
of the construction of superstring field theory
was given in Berkovits' works 
\cite{Berkovits2,Berkovits3}. While previous
string field theories are based on the
Chern-Simons-like actions, Berkovits formulation
is based on the Wess-Zummino-Witten-like
 action. Two major advantages
of Berkovits superstring field theory
 are that it is manifestly
$SO(3,1)$ super-Poincare invariant and
that it does not require contact terms to
remove tree-level divergences
\cite{BerkovitsD} (For recent
review of Berkovits superstring field theory 
(NSFT), see \cite{Ohmori,Desmet,BerkovitsR}.)
In particular, it was shown that the calculation
of the tachyon potential in BSFT is in a
very good agreement with Sen's conjecture
\cite{SenP}.
The tachyon kink and lump solution was
also analysed (
For extensive list of references, see \cite{Ohmori,Desmet}.) 

In this short note we will construct solution
 of NSFT following mainly the seminal
paper \cite{Horowitz}. We begin with
the analysis of the equation of motion of the NS sector
\footnote{In this paper we will discuss Neveu-Schwarz sector
of open string field theory only. However proposal
for inclusion of Ramond sector has been done  in
 very interesting paper \cite{BerkovitsRamond}, see also
\cite{Japan}.}
of NSFT. This string field  theory is  
characterised with  given bulk (bulk
on the world-sheet) conformal field theory (CFT) and
with given boundary conditions on the boundary
of the world-sheet. In other
words, in terminology of \cite{SenV3}, we have a
string field theory characterised  by given boundary CFT (BCFT),
where string fields belong to the Hilbert space of BCFT.
 As is well known  (for recent discussion of this approach in
bosonic open string field theory and 
extensive list of references,
see, for example \cite{SenV1,SenV2,SenV3,Takahashi1},
for case of NSFT see, \cite{KlusonNSFT,KlusonNSFT2,Japan,Marino})
 fluctuations around field configuration  that is  solution
of the equation of motion are described
with the same action as the original one,
however with shifted BRST operator.
Since this new BRST operator can have very complicated form, 
there should exist such
a field redefinition so that after performing this redefinition
in the action, the new BRST operator gains the same structure
as the original one, but fluctuation fields are transformed to
the fields that belong to the Hilbert space of the new BCFT'.
This new BCFT' has the same CFT in the bulk of the world-sheet
as BCFT has 
so that the BRST operator should be the same as the original one,
however boundary conditions have changed which is reflected in
the change of vertex operators and hence string fields. 
 We will show that the generator of this
transformation is in natural way related to the solution of
the equation of motion of NSFT so that whenever we can
find such a generator,  we can very easily find exact solution
of the string field equation of motion. We hope that our
results could be helpful for finding new exact solutions of
the string field theory equation of motion and could be also
helpful in present study of the vacuum string field theory
\cite{SenV1,SenV2,SenV3} (For recent works and extensive
list of references, see \cite{SenV5,Ohmori1}).

This note is organised as follows. In section (\ref{second})
we briefly review basic facts about
NS sector of Berkovits string field theory.

The main part of this paper is contained in section (\ref{third}).
There we will show that for any solution of the equation of
motion of NSFT we can construct  new BRST operator
according to the analysis \cite{KlusonNSFT,Marino}. Using particular
form of this operator suggested in \cite{BerkovitsBRST} we
will show that there exist field redefinition that maps the new 
BRST operator to the original one. Under this transformation
string fields that took values in Hilbert space of the original
BCFT are mapped to the new BCFT' with new boundary conditions.
Then we show that the action is invariant under this redefinition.
Finally, we extend our analysis to the case of Witten's
bosonic open string field theory.

In conclusion (\ref{fifth}) we outline our results and
suggest possible directions of further research.

\section{Review of superstring field theory}
\label{second}
 
In this section we would like to review basic facts about
superstring field theory, for more details, see \cite{Ohmori,
Berkovits1,Berkovits2,Sen1}.
The general off-shell string field configuration
in the GSO(+) NS sector corresponding Grassmann
even open string vertex operator $\Phi$ of ghost number
$0$ and picture number $0$ in the combined
conformal field theory of a $c=15$ superconformal
matter system, and $b,c,\beta,\gamma$ ghost
system with $c=-15$. We can also express $\beta,
\gamma$ in terms of ghost fields
$\xi,\eta,\phi$
\begin{equation}
\beta=\partial \xi e^{-\phi}, \
\gamma=\eta e^{\phi} \ ,
\end{equation}
the ghost number $n_g$ and the picture number $n_p$
assignments are as follows
\begin{eqnarray}
b: \ n_g=-1,\ n_p=0 \ \ \ c: \ n_g=1, \ n_p=0 \ ; \nonumber \\
e^{q\phi}: \  n_g=0, \ n_p=q \ ; \nonumber \\
\xi: \ n_q=-1 , \ n_p=1 \ \ \ \eta: \ n_q=1, \ n_p=-1 \ . \nonumber \\
\end{eqnarray}
The BRST operator $Q_B$ is given
\begin{equation}
Q_B=\oint dz j(z)=\int dz
\left\{c(T_m+T_{\xi\eta}+T_{\phi})+c\partial cb+\eta e^{\phi} G_m
-\eta \partial \eta e^{2\phi}b\right\} \ ,
\end{equation}
where
\begin{equation}
T_{\xi\eta}=\partial \xi \eta, \
T_{\phi}=-\frac{1}{2}\partial \phi\partial\phi-\partial^2\phi \ ,
\end{equation}
$T_{m}$ is a matter stress tensor and $G_m$ is a matter superconformal 
generator. Throughout this paper we will be working in
units $\alpha'=1$.

The string field action is given \cite{Berkovits1,Berkovits2}
\begin{equation}\label{NSFTaction}
S=\frac{1}{2}\int\left(
(e^{-\Phi}Q_Be^{\Phi})(e^{-\Phi}\eta_0
e^{\Phi})-\int_0^1 dt
(e^{-t\Phi}\partial_te^{t\Phi})\left\{
(e^{-t\Phi}Q_Be^{t\Phi}),
(e^{-t\Phi}\eta_0e^{t\Phi})\right\}\right) \ ,
\end{equation}
where $\{A,B\}=AB+BA$ and $e^{-t\Phi}\partial_t
e^{t\Phi}=\Phi$. Here the products and
integral are defined by Witten's gluing prescription
of the string. The exponential of string field is
defined in the same manner $e^{\Phi}=
1+\Phi+\frac{1}{2}\Phi\star\Phi+\dots$. In the
following we will not explicitly write $\star$ symbol.
The basis properties of $Q_B,\eta_0$ which we
will need in our analysis (for more details, see
\cite{Ohmori} and reference therein) are
\begin{eqnarray}\label{ax}
Q_B^2=0, \ \eta_0^2=0, \ \{Q_B,\eta_0\}=0 \ ,
\nonumber \\
Q_B(\Phi_1\Phi_2)=Q_B(\Phi_1)\Phi_2+
\Phi_1Q_B(\Phi_2) , \ \nonumber \\
\eta_0(\Phi_1\Phi_2)=\eta_0(\Phi_1)\Phi_2+
\Phi_1\eta_0(\Phi_2) , \ \nonumber \\
\int Q_B(\dots)=0 \ , \int \eta_0(\dots)=0 \ , \nonumber \\
\end{eqnarray}
where $\Phi_1,\Phi_2$ are Grassmann even fields.

\section{General solution
 in
Berkovits string field theory}\label{third}

Since we will not perform any explicit
calculation we will again use abstract
Witten's formalism in string field theory
\cite{WittenSFT}. As usual we will not
write explicitly the string field theory
star product $\star$.  
We start with   the action 
\begin{equation}\label{VNSFT}
S=\frac{1}{2}\int\left(
(e^{-\Phi}Q_Be^{\Phi})(e^{-\Phi}\eta_0
e^{\Phi})-\int_0^1 dt
(e^{-t\Phi}\partial_te^{t\Phi})\left\{
(e^{-t\Phi}Q_Be^{t\Phi}),
(e^{-t\Phi}\eta_0e^{t\Phi})\right\}\right) \ .
\end{equation}
We must stress that this action is formulated
around one fixed BCFT with given
BRST operator $Q_B$. 
Let us consider general solution of the 
equation of motion that arises from
the action (\ref{VNSFT})
\begin{equation}\label{eqm1}
\eta_0(G_0^{-1}Q_B(G_0))=0 \ , \ G_0=e^{\Phi_0} \  \ .
\end{equation}
Now we would like to study the fluctuation
around this  solution. For that reason
we write general string field containing 
fluctuation around this solution as
\begin{equation}
G=G_0h, \  h=e^{\phi}, \ G^{-1}=h^{-1}G^{-1}_0 \ . 
\end{equation}
To see  that this  field really describes
 fluctuations around  solution $G_0$ note
that for $\phi=0, G=G_0$. It is also clear that
any  string field in the form $e^{\Phi_0+\phi'}$ can be
always rewritten in the form  given above.

  Inserting this upper 
expression in (\ref{VNSFT}) we obtain an action for
$\phi$. As was argued in \cite{KlusonNSFT} in order 
to find form of the new BRST operator
we must ask an question what form
of the equation of motion obeys shifted field $h=e^{\phi}$
\footnote{For alternative and more elegant approach to
this problem, see \cite{Marino}.}.
 Then it was shown \cite{KlusonNSFT} that the new BRST operator has 
a form
\begin{equation}\label{newBRST}
\tilde{Q}(X)=Q_B(X)+A X-(-1)^XX A \ ,
A=G_0^{-1} Q_B(G_0) \ .
\end{equation}
Let us now presume that the new BRST operator
$\tilde{Q}$ can be written as
\footnote{This suggestion is based on results 
presented in \cite{BerkovitsBRST}.}
\begin{equation}\label{newBRSTK}
\tilde{Q}=e^{-K}Q_B(e^{K}) \ ,
\end{equation}
where $K$ is  operator of ghost charge and
picture number equal to zero
that obeys following rules
 \cite{SenV2}
\begin{eqnarray}\label{Krule}
K(A B)=K(A) B+
A K(B) , \ \nonumber \\
\int K(A) B=-\int A  
K(B) \ 
\nonumber \\
\end{eqnarray}
and commutes with $\eta_0$.   
In order to have a well defined string field theory, 
$\tilde{Q}$ must obey (\ref{ax}). Firstly, we can
easily show that
\begin{eqnarray}
\tilde{Q}^2=e^{-K}Q_Be^{K}
e^{-K}Q_Be^{K}=e^{-K}Q^2_B
e^{K}=0 \ , \nonumber \\
\{ \tilde{Q},\eta_0\}=\{
e^{-K}Q_Be^{K},\eta_0\}
=e^{-K}\{Q_B,\eta_0\}e^{K}=0 \ ,
\nonumber \\ 
\end{eqnarray}
since $[K,\eta_0]=0$. We also have
\begin{eqnarray}
\tilde{Q}(\Phi_1\Phi_2)=
e^{-K}\left[Q_B\left(e^{K}(\Phi_1\Phi_2)\right)\right]=
e^{-K}\left[Q_B\left(e^{K}(\Phi_1)e^{K}(\Phi_2)\right)
\right]=
\nonumber \\
=e^{-K}\left[Q_B\left(e^{K}(\Phi_1)\right)e^{K}(\Phi_2)+
e^{K}(\Phi_1)Q_B\left(e^{K}(\Phi_2)\right)\right]=\nonumber \\
=e^{-K}\left[Q_B\left(e^{K}(\Phi_1)\right)\right]
e^{-K}e^{K}(\Phi_2)+e^{-K}
e^{K}(\Phi_1)e^{-K}\left[Q_B\left(e^{K}(\Phi_2)\right)\right]=
\nonumber \\
=\tilde{Q}(\Phi_1)\Phi_2+\Phi_1\tilde{Q}(\Phi_2) \ , \nonumber \\
\end{eqnarray}
where we have used
\begin{eqnarray}
e^{K}(A B)=
A B+ K(A) B+
AK(B)+ \nonumber \\
+ \frac{1}{2}K^2(A) B+ K(A)
 K(B)+ \frac{1}{2}A K^2(B)+
\dots =e^{K}(A) e^{K}(B)
\nonumber \\
\end{eqnarray}
which follows from (\ref{Krule}). In the same way we can show
that
\begin{equation}
e^{-K}(AB)=e^{-K}(A)e^{-K}(B) \ .
\end{equation}
 We can also easily see
that (\ref{newBRSTK}) obeys the last axiom
in (\ref{ax}) 
\begin{equation}\label{Q0}
\int \tilde{Q}(X)=
\int  e^{-K}\left[Q_B\left(e^{K}(X)\right)\right]\mathcal{I}=
\int Q_B\left(e^{K}(X)\right)e^K(\mathcal{I})=
\int Q_B\left(e^{K}(X)\right)=0 
 \ ,
\end{equation}
where $\mathcal{I}$ is "identity" ghost number and
picture number zero field (for recent discussion of
some properties of this field, see \cite{Feng1,Ohmori1})
which is defined as
\begin{equation}
\mathcal{I} X=X \mathcal{I}=X \ , \forall X \ .
\end{equation}
Then we  have 
\begin{equation}
K(X)=K(\mathcal{I}X)=
K(\mathcal{I})X+\mathcal{I}K(X)
=K(X) \Rightarrow K(\mathcal{I})=0
\end{equation}
and consequently 
\begin{equation}
e^{-K}(\mathcal{I})=
e^K(\mathcal{I})=\mathcal{I} \ .
\end{equation} 
In (\ref{Q0}) we have also used
\begin{eqnarray}
\int e^{-K}(X)Y=
\int \sum_n \frac{1}{n!}(-1)^n K^n(X)Y=\nonumber \\
=\sum_n -\int \frac{1}{n!}(-1)^{n}K^{n-1}(X)K(Y)=
\sum_n   \frac{1}{n!} \int X K^n(Y)=
\int X e^{K}(Y)  \nonumber \\
\end{eqnarray}
which is a result of successive application of the
second property of $K$ given  in (\ref{Krule}).

To finish our analysis, we must find such a string field that
solves the equation of motion (\ref{eqm1}) and
leads to the shifted BRST operator (\ref{newBRST}).
In fact, similar calculation has been performed in
 \cite{KlusonNSFT2} so that we do not 
repeat it here. In the same way as in
\cite{KlusonNSFT2} one can show that the field
$\Phi_0$ that is a solution of (\ref{eqm1}) and
leads to (\ref{newBRST}) has a form
\begin{equation}
\Phi_0=K^L(\mathcal{I}) \ ,
\end{equation}
where indices $R,L$ corresponds to the left and right
side of open string \cite{WittenSFT,Horowitz}.

We have seen that the new operator $\tilde{Q}$ obeys
all axioms given in (\ref{ax}) and hence we can
define NSFT around the new background configuration
characterised with this new BRST operator
 (\ref{newBRST}). It is important
to stress that in BCFT language, we have  new BRST
operator that can be completely general, however 
the fluctuation fields still belong to the Hilbert space
of the original BCFT. For that reason it is natural to
find such a field redefinition that maps $\tilde{Q}$ in
$Q_B$ so that the new BRST operator is
 of the same form and the fluctuation
fields that belong to the Hilbert space of the original
BCFT are mapped to the fields that belong to the Hilbert space of
new BCFT'. We will show that this can be easily done
in the framework of Berkovits string field theory.

In order to obtain BRST operator of
the same functional form as the original one,
 we perform field redefinition
\begin{equation}
\phi=e^{-K}(\Psi) \ .
\end{equation}
Using  (\ref{Krule}) we
can  show 
\begin{eqnarray}
e^{\phi}=\sum_{n=0}^{\infty}\frac{1}{n!}
\left(e^{-K}(\Psi)\right)^n=
e^{-K}\left(\sum_{n=0}^{\infty}\frac{1}{n!}
\Psi^n\right)=e^{-K}(e^{\Psi}) \ , \nonumber \\
e^{-\phi}=\sum_{n=0}^{\infty}
\frac{(-1)^n}{n!}\left(e^{-K}(\Psi)\right)^n
=e^{-K}\left(\sum_{n=0}^{\infty}
\frac{(-1)^n}{n!}\Psi^n\right)=
e^{-K}(e^{-\Psi}) \  \nonumber \\
\end{eqnarray}
and consequently
\begin{equation}
e^{-K}(e^{\Psi})e^{-K}(
e^{-\Psi})=e^{-K}(e^{\Psi}
e^{-\Psi})=e^{-K}(\mathcal{I})
=\mathcal{I} \ .
\end{equation}
Using these results we immediately obtain
\begin{eqnarray}\label{first}
\int (e^{-\phi}
\tilde{Q}(e^{\phi}))(e^{-\phi}\eta_0(e^{\phi}))=
\nonumber \\ =
\int e^{-K}(e^{-\Psi})e^{-K}
Q_Be^{K}e^{-K}(e^{\Psi})
e^{-K}(e^{-\Psi})
\eta_0(e^{-K}(e^{-\Psi}))= \nonumber \\
=\int  e^{-\Psi}e^{K}\left[
e^{-K}Q_B(e^{\Psi})
e^{-K}(e^{-\Psi})\eta_0(e^{-K}(e^{\Psi}))
\right]=\nonumber \\ =
\int e^{-\Psi}e^{K}\left[
e^{-K}(Q_B(e^{\Psi}))e^{-K}(
e^{-\Psi}\eta_0(e^{\Psi}))\right]=
\nonumber \\
=\int (e^{-\Psi}Q_B(e^{\Psi}))(e^{-\Psi}\eta_0(
e^{\Psi})) \ . \nonumber \\
\end{eqnarray}
We can also show that
\begin{eqnarray}
e^{t\phi}=\sum_n \frac{1}{n!}t^n
(e^{-K}(\Psi))^n=
e^{-K}\left(\sum_n \frac{1}{n!}
(t\Psi)^n\right)=e^{-K}(e^{t\Psi})  \ , \nonumber \\
e^{-t\phi}\partial_t e^{t\phi}=
e^{-K}e^{-t\Psi}e^{-K}
\partial_t(e^{t\Psi})=e^{-K}(e^{-t\Psi}
\partial_t e^{t\Psi}) \ , \nonumber \\
e^{-t\phi}\tilde{Q}e^{t\phi}=
e^{-K}(e^{-t\Psi})e^{-K}
Q_Be^{K}e^{-K}(e^{t\Psi})
=e^{-K}\left(e^{-t\Psi}Q_B(e^{t\Psi})\right) , \ \nonumber \\
e^{-t\phi}\eta_0( e^{t\phi})=e^{-K}
\left (e^{-t\Psi}\eta_0(e^{t\Psi})\right) \nonumber \\
\end{eqnarray}
so that we have
\begin{eqnarray}\label{sec}
\int_0^1 dt
(e^{-t\phi}\partial_te^{t\phi})\left\{
(e^{-t\phi}\tilde{Q}(e^{t\phi})),
(e^{-t\phi}\eta_0(e^{t\phi}))\right\}=\nonumber \\=
\int_0^1 dt
e^{-K}(e^{-t\Psi}\partial_te^{t\Psi})e^{-K}
\left\{
(e^{-t\Psi}Q_B(e^{t\Psi})),
(e^{-t\Psi}\eta_0(e^{t\Psi}))\right\}= \nonumber \\
=\int_0^1 dt
(e^{-t\Psi}\partial_te^{t\Psi})e^{K}
e^{-K}\left\{
(e^{-t\Psi}Q_B(e^{t\Psi})),
(e^{-t\Psi}\eta_0(e^{t\Psi}))\right\}=\nonumber \\
=\int_0^1 dt
e^{-t\Psi}\partial_te^{t\Psi}\left\{
(e^{-t\Psi}Q_B(e^{t\Psi})),
(e^{-t\Psi}\eta_0(e^{t\Psi}))\right\} \ .
\nonumber \\
\end{eqnarray}
From (\ref{first}), (\ref{sec}) we obtain correct form of
the action for fluctuation fields  around new background
configuration, however with the BRST operator having
the same form as the original one. Now the modification
of the background is reflected in the change of
the string fields that now belong to the Hilbert space of
the BCFT' theory. 

We can  easily previous result  to the
case of 
Witten's open bosonic string theory with the action
\begin{equation}\label{Wittenaction}
S=\int\frac{1}{2}\Phi Q_B(\Phi)+
\frac{1}{3}\Phi \Phi  \Phi \ ,
\end{equation}
where now $\Phi$ is ghost number one string field.
From (\ref{Wittenaction}) we obtain the
 equation of motion
\begin{equation}
Q_B\Phi+\Phi\Phi=0 \ .
\end{equation}
It is easy  to see that the field
\begin{equation}
\Phi_0=e^{-\Psi_0}Q_B(e^{\Psi_0})
\end{equation}
with any ghost number zero field
$\Psi_0$ is  a solution of the equation of motion.
As in previous case we presume that  $\Psi_0$ 
can be written  as follows 
\begin{equation}
\Psi_0=K_L(\mathcal{I}) \ ,
\end{equation}
where $K$ is  operator obeying (\ref{Krule}).
Now the new BRST operator has the same
form as before
\begin{equation}
e^{-K}Q_B(e^{K}) \ .
\end{equation}
 Then we can perform
the same field transformation as in supersymmetric
case and we end up with the action,   where the
BRST operator has the  same form as the original one
and the fluctuating field belongs to the Hilbert 
space of the new BCFT' theory. So that whenever
we will find such  two BCFT theories that are related
field redefinition given above, we can claim that one
string field theory action arises from the expansion
around the background configuration that is
solution of the equation of motion of
the original string field theory where the solution 
can be written as $\Phi_0=K^L(\mathcal{I})$. 

\section{Conclusion}\label{fifth}
In this short note we gave a simple solution of
the equation of motion of the Berkovits string field theory.
 We have shown that whenever we
can express shifted BRST operator in the form similar to
the form suggested in
\cite{BerkovitsBRST} we can easily obtain the solution of
the equation of motion. On the other hand, knowledge of
the  operator $K$ allows us to find such a solution
as well. In particular, operator $K$ can correspond to the
generator of broken symmetry by D-brane background as for
example, generator of translation in direction transverse to
D-brane. Then solution given above corresponds to translation
of D-brane in transverse direction. It is clear that these two
configurations should be equivalent so that the original and
final BCFT should be the same. We will show in forthcoming
publication that this is really true.

We have also shown that similar method can be used in
case of Witten's bosonic string field theory. However,
in this case the appropriate solution corresponds to the
pure gauge.  The similar result has been discussed recently
in \cite{Takahashi1,Takahashi2}. 

Finally we must stress that our method cannot generate all solutions
of equation of motion of string field theory. From discussion given above
it is clear that this method is not applicable when we cannot
find operator $K$ relating original and final configuration. In particular,
this method cannot be used for the case of tachyon condensation. 

There are many problems that worth to be studied.
Firstly, we will present particular examples, corresponding
to the marginal deformation of the original string field theory.
Then we will try to formulate our result in the CFT language,
using very elegant formalism reviewed recently
in \cite{SenV5,Ohmori1}. We will return to these problems in
future. And finally, we hope to extend our result to the case
of NSFT theory including Ramond sector, following recent
papers \cite{BerkovitsRamond,Japan}.
\\
\\
{\bf Acknowledgement}

This work is partly supported by EU contract
HPRN-CT-2000-00122.


\end{document}